\documentclass[journal]{IEEEtran}
\usepackage{booktabs}
\usepackage{makecell}
\usepackage{graphicx}
\usepackage[table]{xcolor} 
\usepackage{amsmath}
\usepackage{hyperref}
\usepackage{pgfplots}
\pgfplotsset{compat=newest}
\usepackage{subfig}
\usepackage{caption}

\usepackage{hyperref}

\ifCLASSINFOpdf
\else
\fi
\usepackage{amsmath}
\usepackage{array}

\usepackage{stfloats}
\begin{document}
%
\title{Quantitative and Qualitative Evaluation of NLM and Wavelet Methods in Image Enhancement}
%
%
%

\author{
    \IEEEauthorblockN{Cameron Khanpour}
    \\
    \IEEEauthorblockA{\textit{Georgia Institute of Technology}}
}

\maketitle

\begin{abstract}
This paper presents a comprehensive analysis of image denoising techniques, primarily focusing on Non-local Means (NLM) and Daubechies Soft Wavelet Thresholding, and their efficacy across various datasets. These methods are applied to the CURE-OR, CURE-TSD, CURE-TSR, SSID, and Set-12 datasets, followed by an evaluation using Image Quality Assessment (IQA) metrics PSNR, SSIM, CW-SSIM, UNIQUE, MS-UNIQUE, CSV, and SUMMER. The results indicate that NLM and Wavelet Thresholding perform optimally on Set12 and SIDD datasets, attributed to their ability to effectively handle general additive and multiplicative noise masks. However, their performance on CURE datasets is limited due to the presence of complex distortions like Dirty Lens and Codec Error, which these methods are not well-suited to address. Analysis between NLM and Wavelet Thresholding shows that while NLM generally offers superior visual quality, Wavelet Thresholding excels in specific IQA metrics, particularly SUMMER, due to its enhancement in the frequency domain as opposed to NLM's spatial domain approach.
\end{abstract}


%
\IEEEpeerreviewmaketitle

\section{Introduction}
%
%
%
%
\IEEEPARstart{I}{n} any image, noise and distortions may be present and affect the clarity of the image. One of the biggest drivers in the field of Digital Image Processing is the process of understanding the noise and distortions of the image, denoising and enhancing the image, then quantifying those results with Image Quality Assessment (IQA) metrics that can either mock the Human Visual System (HVS) or tell more about the results of the denoising process. Denoising images is also a crucial step for computer visual systems, allowing the application of image denoising and enhancement to extend to object detection in challenging scenarios. 
\par This paper will utilize two image denoising methods, Non-local Means (NLM) and Daubechies Soft Wavelet Thresholding, on the CURE-OR, CURE-TSD, CURE-TSR, SSID, and Set-12 datasets. The resulting images will be evaluated using IQA metrics such as PSNR, SSIM, CW-SSIM, UNIQUE, MS-UNIQUE, CSV, and SUMMER. The paper will then go into further analysis of the different types of noise and distortions in each dataset and showcase what parameters can be adjusted to improve that denoising method, as well as showcase what types of distortions that denoising method best enhances.

\section{Background} 
\subsection{Traditional Method for Image Denoising}
Classical assumptions used to understand the distortions in an image include assuming that neighboring pixels in an image are highly correlated. Therefore, filtering the image by averaging nearby pixels via Gaussian Smoothing is preformed to denoise an image. More mathematically,
\begin{equation}
\hat{f}[n] = \frac{1}{C_n} \sum_{k} g(k)e^{-\frac{\|n-k\|^2}{2\sigma^2}}
\end{equation}
where \(\hat{f}[n]\) is a denoised image from a noisy image \(g(k)\) that iterates through a kernel where \(\sigma^2\) is the standard deviation of the image. This is effectively a low pass filter in the frequency domain that creates a uniform blur across the image, eliminating a large portion of the noise at the trade off of losing the fine details of the original image. This paper extends this method utilizing NLM and Wavelet Threshold denoising.

\subsection{IQA Metrics}
Two popular objective image quality metrics are PSNR and SSIM. PSNR between two images \( x \) and \( y \) can be calculated as follows:

\begin{equation}
PSNR = 10 \log_{10} \left( \frac{\max(I_{\text{original}})^2}{MSE_{xy}} \right).
\end{equation}

The SSIM is given by:

\begin{equation}
SSIM = \frac{(2\mu_x\mu_y + c_1)(2\sigma_{xy} + c_2)}{(\mu_x^2 + \mu_y^2 + c_1)(\sigma_x^2 + \sigma_y^2 + c_2)}.
\end{equation}
where \( \mu_x \) and \( \sigma_x \) are the mean and standard deviation of image \( \alpha = x, y \), and \( \sigma_{xy} \) is the covariance of \( x \) and \( y \). \( c_i = (k_iL)^2 \), \( i = 1, 2 \) with \( k_1 = 0.01 \), \( k_2 = 0.03 \), and \( L \) being 255 [1].

CW-SSIM (Complex Wavelet Structural Similarity Index) modifies SSIM for application in the complex wavelet domain, providing robustness to small geometric distortions.

\begin{equation}
\text{CW-SSIM} = \frac{2 \sum\limits_{l=1}^{L} \!\left| \sum\limits_{k \in \Omega} \!x_{kl} y_{kl}^* \right| + K}{\sum\limits_{l=1}^{L} \!\left( \sum\limits_{k \in \Omega} \!|x_{kl}|^2 + \sum\limits_{k \in \Omega} \!|y_{kl}|^2 \right) + K}.
\end{equation}
where \(K\) is a small positive constant to improve the robustness of the CW-SSIM measure when the local signal to noise ratios are small [2].

The CSV (Color Structure and Visual system) metric is a full-reference estimator that quantifies perceptual color degradations, structural, and perceptual differences. It utilizes \( CIEDE \) to  represent the color difference based on CIEDE2000, Color Name Distance, Structural Difference, and Retinal Ganglion Cell-based Difference to closely match to HVS perceived fidelity [3].

UNIQUE (Unsupervised Image Quality Estimation) uses sparse representations obtained through unsupervised learning. Its quality estimation is based on the Spearman rank order correlation coefficient of sparse representations [4].

MS-UNIQUE extends UNIQUE using multiple linear decoders. Its quality estimation involves a weighted combination of filter responses [5].

SUMMER (SUM of Modified Error Ratios) assesses image quality by emphasizing perceptual features in the Fourier domain by using location information from analyzing the magnitude spectrums over each color channel and use frequency-based weights to align quality scores [6].

\section{Method}
\subsection{Non-Local Means (NLM)}
The idea of NLM is a weighted average of all pixels in the image to get a denoised image \( z \):
\begin{equation}
z = \frac{\sum_{j} \omega_{ij} y_j}{\sum_{j} \omega_{ij}}
\end{equation}
and \( \omega_{ij} \) is the weights that measures the similarity between the neighborhood around the \( i \)th pixel and the \( j \)th pixel:
\begin{equation}
\omega_{ij} = \frac{1}{Z_i} \exp\left(-\frac{\sum_{m} k_m \left[v(y_i)_m - v(y_j)_m\right]^2}{h^2}\right)
\end{equation}
where \( Z_i \) is defined as:
\begin{equation}
Z_i = \sum_{j} \exp\left(-\frac{\sum_{m} k_m \left[v(y_i)_m - v(y_j)_m\right]^2}{h^2}\right)
\end{equation}
Here \( v(y_i) \) is the neighborhood around the \( i \)th pixel in the image, \( m \) denotes the pixels in the neighborhood, \( k_m \) is a Gaussian kernel, and \( h \) controls the decay of the weight [1].

The \texttt{denoise\_nl\_means()} function in the \texttt{skimage} module of Python was used. The parameter \( h \) was chosen to be \( 0.8 \times \sigma_{\text{est}} \), where \( \sigma_{\text{est}} \) is the estimated standard deviation of Gaussian Noise in the image. The constant 0.8 was multiplied to \( \sigma_{\text{est}} \) to allow less blurring of fine details of the denoised image. During later analysis, it will be shown how changing this constant can improve denoising of different kinds of distortions.

\subsection{Wavelet Thresholding}
Wavelet Thresholding is a method for image denoising that exploits the multi-resolution representation of wavelet transforms. The \texttt{denoise\_wavelet()} function from the \texttt{skimage} module in Python implements this technique. The approach involves decomposing an image into a set of wavelet coefficients, applying thresholding to these coefficients, and then reconstructing the image from the modified coefficients. The process can be described as follows:

Given a noisy image \( y \), the wavelet transform decomposes it into a set of coefficients \( W(y) \), which includes both approximation coefficients (low-frequency components) and detail coefficients (high-frequency components). Soft thresholding is then applied to modify these coefficients according to the function:

\begin{equation}
\hat{w}_{ij} = 
\begin{cases} 
\text{sgn}(\omega_{ij}) \cdot (|\omega_{ij}| - \lambda), & \text{if } |\omega_{ij}| \geq \lambda, \\
0, & \text{if } |\omega_{ij}| < \lambda.
\end{cases}
\end{equation}
where \( \hat{w}_{ij} \) are the thresholded wavelet coefficients, \( \omega_{ij} \) are the original wavelet coefficients, and \( \lambda \) is the threshold value, which is chosen based on the estimated noise standard deviation \( \sigma_{\text{est}} \). This soft thresholding process attenuates the coefficients by reducing their magnitude, thus reducing noise and preserving significant image features [7]. The denoised image \( z \) is then reconstructed from these threshold coefficients by taking the inverse of the transform.

Daubechies 'db2' wavelet was used for its good balance between performance and computational efficiency. The \texttt{denoise\_wavelet()} function automatically sets \( \lambda \) and \( \sigma_{\text{est}} \) is calculated similarly to NLM. The choice of soft thresholding ensures a more natural denoising effect by preserving the continuity of the wavelet coefficients, which is particularly effective in maintaining the sharpness and textures of the original image.

\section{Results}
Utilizing NLM and Wavelet Thresholding, images from the Set12, SIDD, CURE-OR, CURE-TSD, and CURE-TSR were denoised [8-10]. The grayscale Set12 image set was given three levels of noise of varying difficulty. After each dataset was denoised, IQA metrics were ran and averaged across all of the denoised images for each dataset.

\begin{table}[htbp]
\centering
\caption{Average performance of NLM enhancement}
\begin{tabular}{lccccc}
\toprule
\makecell{IQA\\Metrics} & Set12 & SIDD & CURE-OR & CURE-TSD & CURE-TSR \\
\midrule
PSNR (dB)& 32.190 & 34.509 & 25.257 & 23.956 & 23.644 \\
SSIM     & 0.7620 & 0.7795 & 0.4782 & 0.4473 & 0.4357 \\
CW-SSIM  & 0.9314 & 0.7492 & 0.5452 & 0.5395 & 0.5283 \\
UNIQUE   & 0.6464 & 0.8763 & 0.1496 & 0.1438 & 0.1395 \\
MS-UNIQUE& 0.7602 & 0.9128 & 0.4124 & 0.3821 & 0.3913 \\
SUMMER   & 3.6747 & 3.4506 & 0.7162 & 0.6040 & 0.6150 \\
CSV      & 0.9803 & 0.9756 & 0.9547 & 0.8490 & 0.7931 \\
\bottomrule
\end{tabular}
\label{tab:nlm_performance}
\end{table}

\begin{table}[htbp]
\centering
\caption{Average performance of Wavelet Thresholding enhancement}
\begin{tabular}{lccccc}
\toprule
\makecell{IQA\\Metrics} & Set12 & SIDD & CURE-OR & CURE-TSD & CURE-TSR \\
\midrule
PSNR (dB)& 31.452 & 33.774 & 25.884 & 23.752 & 23.363 \\
SSIM     & 0.6901 & 0.7436 & 0.3791 & 0.3642 & 0.3588 \\
CW-SSIM  & 0.9261 & 0.7157 & 0.6232 & 0.5981 & 0.5877 \\
UNIQUE   & 0.5572 & 0.8690 & 0.2039 & 0.1895 & 0.1911 \\
MS-UNIQUE& 0.7091 & 0.9148 & 0.2712 & 0.2680 & 0.2708 \\
SUMMER   & 3.8866 & 3.5701 & 2.2341 & 1.9863 & 2.0412 \\
CSV      & 0.9802 & 0.9769 & 0.9428 & 0.8691 & 0.8703 \\
\bottomrule
\end{tabular}
\label{tab:wavelet_performance}
\end{table}

As shown in the tables above, NLM and Wavelet Thresholding preformed significantly better across all IQA metrics in the Set12 and SIDD image sets over the CURE image sets. 

\begin{figure*}[!t]
\centering
\subfloat[Noise Level 15\%]{
\includegraphics[width=0.3\textwidth]{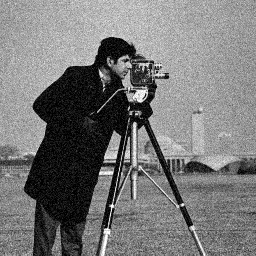}
\label{fig:noise_15}
}
\hfil
\subfloat[Noise Level 35\%]{
\includegraphics[width=0.3\textwidth]{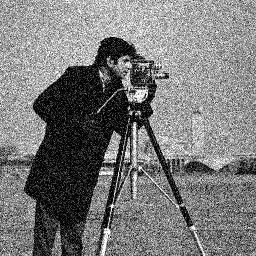}
\label{fig:noise_35}
}
\hfil
\subfloat[Noise Level 50\%]{
\includegraphics[width=0.3\textwidth]{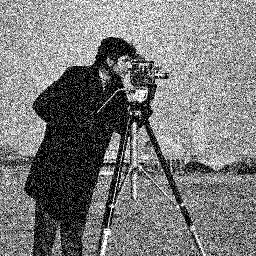}
\label{fig:noise_50}
}
\hfil
\subfloat[Noise Level 15\% Denoised NLM]{
\includegraphics[width=0.3\textwidth]{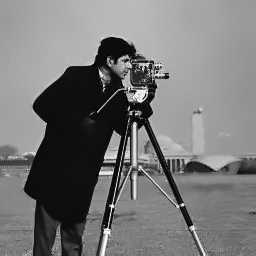}
\label{fig:denoised_15_nlm}
}
\hfil
\subfloat[Noise Level 35\% Denoised NLM]{
\includegraphics[width=0.3\textwidth]{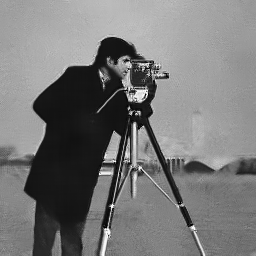}
\label{fig:denoised_35_nlm}
}
\hfil
\subfloat[Noise Level 50\% Denoised NLM]{
\includegraphics[width=0.3\textwidth]{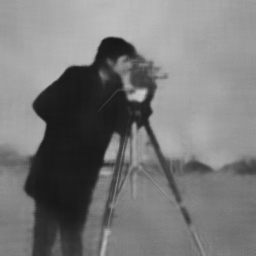}
\label{fig:denoised_50_nlm}
}

\caption{Comparison of noisy images at different levels and their denoised versions using NLM.}
\label{fig:images}
\end{figure*}

Another interesting result is the higher score in the SUMMER metric's average values for Wavelet Thresholding compared to NLM. This likely is due to SUMMER being an IQA metric based on analyzing the spectrum of the image, and Wavelet Thresholding being a frequency domain based image enhancement gave it better scores compared to the spatial domain based image enhancement NLM.

\section{Analysis}
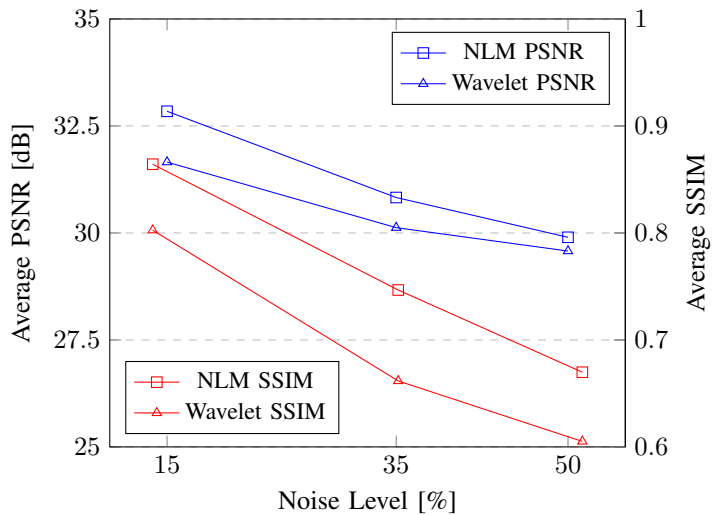
\begin{figure}[htbp]
\centering
\hspace{-.75cm}
\begin{tikzpicture}
\begin{axis}[
    xlabel={Noise Level [\%]},
    ylabel={Average PSNR [dB]},
    xmin=10, xmax=55,
    ymin=25, ymax=35,
    xtick={15,35,50},
    ytick={25,27.5,30,32.5,35},
    legend pos=north east,
    ymajorgrids=true,
    grid style=dashed,
    legend style={font=\small},
]

\addplot[
    color=blue,
    mark=square,
    ]
    coordinates {
    (15,32.84299886566928)(35,30.82838983242492)(50,29.898492636604264)
    };
    \addlegendentry{NLM PSNR}

\addplot[
    color=blue,
    mark=triangle,
    ]
    coordinates {
    (15,31.655584299066177)(35,30.123935097229563)(50,29.577303462290157)
    };
    \addlegendentry{Wavelet PSNR}

\end{axis}

\begin{axis}[
    axis y line*=right,
    axis x line=none,
    ylabel={Average SSIM},
    ymin=0.6, ymax=1,
    ytick={0.6,0.7,0.8,0.9,1.0},
    legend pos=south west,
    ymajorgrids=false,
    grid style=dashed,
    legend style={font=\small},
]

\addplot[
    color=red,
    mark=square,
    ]
    coordinates {
    (15,0.8642579374228368)(35,0.7466570810704711)(50,0.6698272168938863)
    };
    \addlegendentry{NLM SSIM}

\addplot[
    color=red,
    mark=triangle,
    ]
    coordinates {
    (15,0.8025816671356938)(35,0.6616792400988646)(50,0.605135099177592)
    };
    \addlegendentry{Wavelet SSIM}

\end{axis}
\end{tikzpicture}
\caption{Trend of PSNR and SSIM over different levels of noise for Set12.}
\label{fig:trend}
\end{figure}

Figure~\ref{fig:images} shows an image from the Set12 dataset with three varying levels of noise being enhanced via NLM. NLM does a good job, even under heavy noise, at removing distortions from the image. At the higher levels of noise, there is a "smudging" effect on the image. In terms of IQA metrics, Figure~\ref{fig:trend} shows the PSNR and SSIM of the different levels of noise. While there is less of a drop in PSNR, there is a sharper drop in SSIM, which more closely reflects the loss of fidelity from denoising higher levels of noise. 

\begin{figure}[htbp] 
\centering 
\includegraphics[width=0.7\linewidth]{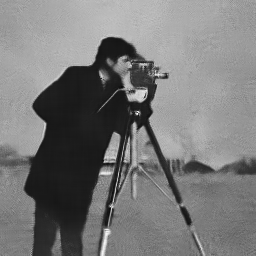} 
\caption{Noise Level 50\% Denoised NLM \(h = 0.6\)} 
\label{fig:denoised_nlm_h=.6} 
\end{figure}

One way to improve the smudging effect from NLM is to change the value of \(h\) to be \( 0.6 \times \sigma_{\text{est}} \) as shown in Figure~\ref{fig:denoised_nlm_h=.6}. By using a lower \(h\) value you can preserve more of the fine details of the image. The cameraman image shows sharper edges along the coat and tripod, as well as more fine details on the face and camera themselves. While lowering \(h\) does help in this case to improve image quality, it does not guarantee improved performance for all images.

Additionally, while this method does improve general additive and multiplicative noise masks, it does not affect the more complex distortions in the CURE datasets.

Comparing the IQA metrics for the SIDD dataset, the brighter images that were denoised did better overall in IQA metrics compared to darker images in the dataset. This is likely due to the fact the noise in SIDD are from smartphone cameras which are limited due to their small aperture and sensor size. So when more light is available to the image, the issues of having a small aperture and sensor size are alleviated, meaning less noise are present in the image. 

In the CURE datasets, NLM and Wavelet Thresholding caused more of a blurring effect in the with the complex distortions like in Dirty Lense. While this slightly alleviated some of the original distortions, the perceptual quality of the images were diminished because of the addition of the blurring. 

Comparing NLM to Wavelet Thresholding directly, IQA metrics show that NLM for the most part surpasses Wavelet Thresholding in image quality (exception being SUMMER). This also matches to visually how the images compare in terms of quality. However, NLM was computationally more intensive than Wavelet Thresholding because Wavelet Thresholding utilized faster algorithms like Fast Wavelet Transform (FWT).

\section{Conclusion}

This paper evaluated the effectiveness of Non-local Means (NLM) and Daubechies Soft Wavelet Thresholding in denoising images across the datasets Set12, SIDD, and the CURE series. The results showed that while both methods excel in reducing general noise in Set12 and SIDD datasets, they struggle with complex distortions like Dirty Lens and Codec Error in the CURE datasets. This highlights a limitation in their adaptability to diverse distortion types.

Comparatively, NLM generally provides better visual quality, but Wavelet Thresholding has a distinct advantage in specific IQA metrics, particularly SUMMER, due to its frequency domain approach. This finding emphasizes the importance of method selection based on noise characteristics and the desired image quality.

The paper also explored parameter tuning, showing that adjustments in NLM can enhance detail preservation but do not universally guarantee improved performance. Additionally, lighting conditions significantly impact noise characteristics, especially in images from smartphone cameras.

Future work could focus on further parameter tuning to more effectively enhance complex distortions found in the CURE datasets. Another potential improvement is in the quality assessment. Object Recognition algorithm could be used on the denoised images of the CURE-TSD and CURE-TSR dataset to show the improvement in traffic sign detection, which has applications in autonomous vehicles.


%

\ifCLASSOPTIONcaptionsoff
  \newpage
\fi

\end{document}